\documentclass[11pt]{revtex4}
\usepackage{mathtools}
\usepackage{hyperref}
\usepackage{pdflscape}
 \usepackage{graphics}
 \usepackage{subfigure}
\usepackage{epsfig}
 \usepackage{amsmath,amsfonts}
 \usepackage{amssymb}
\begin{document}
\title{Gravitational Lensing By A Back  Hole in Poincar{\'e} Gauge Theory of Gravity}
\author{Saboura Zamani}
\email{saboura.zamani@gmail.com}

\author{Siamak Akhshabi}
\email{s.akhshabi@gu.ac.ir}

\affiliation{Department of Physics, Faculty of Sciences, Golestan
University, Gorgan, IRAN }

\begin{abstract}
Using a recently found black hole solution in the framework of the
Poincar{\'e} gauge theory of gravity, we study gravitational lensing for a
system where the lens is a static spherically symmetric black hole. By
analyzing the equations of motion for light rays in a space-time with
torsion, we derive the deflection angle as the light emitted from a source
pass through near the black hole and numerically solve the resulting
integral. We also study the effects of torsion on the position of images. The
results show that the presence of torsion slightly alters both the deflection
angle and position of images in this setup.
\end{abstract}
	
\maketitle
\section{Introduction}
\label{sec: intro}
The discovery of the light bending phenomenon as it passes through a
gravitational field of a celestial object was one of the first and most
important observational tests of Einstein's theory of General Relativity (GR).
This phenomenon, commonly known as gravitational lensing, has various
different applications in astrophysics and cosmology. More specifically, near
a black hole where the gravitational field is extremely powerful,  the path
followed by light rays can  reveal a great deal of information about the
geometry and properties of the surrounding space. On the other hand, this
path as well as the type and shape of any lensing effects are strongly
connected to the background geometry of space-time in which the light is
traversing. The theory of General Relativity, while hugely successful at
scales and energies of solar system tests, is expected to be modified at
extremely high energies and in very strong gravitational fields or at the
scales where quantum effects become important. For these reasons, it is useful
to examine the gravitational lensing in the context of alternative theories
of gravity to determine the necessary corrections to the General Relativistic
results, and these corrections are expected to be more substantial, or at
least observable in a very strong gravitational field of a black hole.

Gravitational lensing also
provides a convenient tool to measure various cosmological parameters. Most
importantly the present day rate of expansion of the universe (\textit{i.e.} the Hubble parameter) can
be measured by the time delay between various images of a variable source. A
theory that alters the results of gravitational lensing, may also provide a
way to resolve the Hubble tension problem \cite{Riess1, Riess2, Bernal, Yang, Wen}.

One of the first comprehensive studies of the strong gravitational lensing by
a Schwarzschild black hole was performed by Virbhadra and Ellis
\cite{Virb1}. In Ref. \cite{Claudel} the geometric structure of the photon
surfaces has been thoroughly studied. Analysis of various characteristics of
gravitational lensing by black holes and naked singularities were also
performed in \cite{Virb2, Virb3, Virb4}. Frittelli \textit{et al.} found a
general exact lens equation independent of the background metric in Ref.
\cite{Frit}. Analytical investigation of the strong black hole lensing was
performed by Bozza \textit{et al.} \cite{Bozza1}  and Bozza \cite{Bozza2, Bozza3} in
the case of a   Schwarzschild background and by Eiroa \textit{et al.} for the
Reissner-Nordstr\"{o}m case \cite{Eiroa}. More recently, these effects have
been analyzed in the framework of various modified gravity theories
\cite{Bhad, Whisk, Tsuj, Rugg, Sot, Sah, Li, Izma, Baha}.

There exist many different proposed modifications to the theory of General
Relativity at high energy scales, each with their own physical justifications
and characteristics. Some of the most natural and important modifications
among them, are gauge theories of gravity which apply the gauge principle
\textit{i.e.} localization of symmetries to the gravitational interaction.
One key outcome of these theories is that the space-time geometry of General
Relativity, Riemannian space-time, is transformed into a non-Riemannian
geometry with curvature and torsion. In these theories, the presence of
torsion, which is coupled to the spin of the matter, can alter the trajectory
of light rays and influence gravitational lensing effects. Also, from a
quantum gravity point of view, many proposed theories for the unification of
quantum mechanics and gravity, include a torsion field one way or another
\cite{Schrek, Hehl1, Hehl2, Shapiro}. As a specific example, in string theory
the effective Lagrangian at low energies has been shown to be equivalent to a
Brans-Dicke generalization of a metric theory of gravity with torsion
\cite{Hammond1, Hammond2}. Moreover, in string theory, there exists a
Kalb-Ramond field whose field strength can act like a torsion field in the
background geometry \cite{Kalb}. The presence of this field is also well
known in noncommutative field theories \cite{Seiberg} where torsion is also
known to be present \cite{Chamseddine}.

On the other hand, the presence of torsion in many gauge theory descriptions
of gravity is also well known. Applying the well-established gauging
procedure (replacing global symmetries with local ones) used in the standard
model of particle physics to the gravitational interaction, one arrives at a
gravitational theory with a general non-symmetric connection, which naturally
incorporates the spin of the matter to the gravitational interactions
\cite{Blag}. The gravitational interactions here are governed by two gauge
potentials, which in a Riemann-Cartan geometry can be interpreted as tetrad
and spin connection fields. The associated field strengths of these gauge
potentials are curvature and torsion tensors \cite{Hayashi}. This theory is
usually called the Poincar{\'e} gauge theory (PGT) of gravity. There exist
some well-known special cases in PGT: General Relativity (vanishing torsion),
teleparallel theory (vanishing curvature) and also Einstein-Cartan theory
which can be regarded as the simplest generalization of General Relativity
and has the same Lagrangian as General Relativity but with a non-symmetric
connection. Since its introduction, the Einstein-Cartan theory has been extensively studied in the literature
\cite{Kerlick}. Here the torsion tensor is related by an algebraic equation
to the spin density of the matter and as a result is not a dynamical
quantity. This means that in Einstein-Cartan theory torsion can not
propagate, \textit{i.e.} there are no gravitational wave modes associated with torsion
\cite{Rauch}. However, by choosing more complicated quadratic Lagrangian in
Poincar{\'e} gauge theory of gravity, propagating torsion modes can be
present and there exist torsion waves in space-time \cite{Blag2}.

The aim of the present paper is to study the effects of non-Riemannian
geometry of the PGT on the deflection angle and position of images for a
gravitational lensing system where the lens is a black hole. In particular, we
examine the effects of torsion and spin on the lensing parameters in this
framework. The structure of the paper is as follows: in Sect. \ref{sec: 2} we
offer a brief introduction to the basic properties of PGT and review a new
static spherically symmetric black hole solution, first derived in Ref.
\cite{Cemb} by analyzing the field equations in this theory. This solution
describes a Reissner-Nordstr\"{o}m type solution where torsion plays a
similar role here to the electric charge in the usual Reissner-Nordstr\"{o}m
geometry of General Relativity. In Sect. \ref{sec: 3}, we study the
gravitational lensing and derive an integral relation for the deflection
angle around a black hole by using both the metric geodesics and
auto-parallel curves in this setup. These integrals are of the elliptic type
and can be solved numerically. The position of images is analyzed using the
black hole lens equation. Finally, Sect. \ref{sec: 4} is dedicated to the
conclusion.

\section{Gauge theories of gravity with torsion}
\label{sec: 2}
As stated above, the geometric structure of PGT is a Riemann-Cartan space-time where curvature and torsion tensor are given in terms of the dynamical variables (tetrad and spin connection) by the following relations
\begin{eqnarray}
    \label{Rim1}
   {R^{j}_{~\mu\nu i}}&=&2\Big({\partial_{[\mu}T^j_{~\nu]i}}+{\Gamma^j_{~~[\mu|k}}
    {\Gamma^k_{~~|\nu]i}}\Big)\;,
\\ \nonumber
\label{Tor1}
{T^i_{~\mu\nu}}&=&2\Big({\partial_{[\mu}e^i_{~\nu]}}+{\Gamma^i_{~[\mu|j}}
{e^j_{~|\nu]}}\Big)\,\,,\,\, T_\mu={T^\nu_{~\mu\nu}}
\end{eqnarray}
where $e_{~\mu}^{i}$ is the tetrad field and
\begin{equation}
g_{\mu\nu}=\eta_{ij}e_{~\mu}^{i}e_{~\nu}^{j}\;,
\end{equation}
is the space-time metric. The relation between the spin connection and  the ordinary affine
connection is given by the equation below
\begin{equation}
\partial_{\mu}e^{i}_{\nu}+\Gamma^i_{~j\mu}e^j_{~\nu}-\Gamma^{\lambda}_{~~\mu\nu}e^i_{~\lambda}=0.
\end{equation}

Throughout the paper, the Greek indices will refer to the holonomic coordinate
bases of the manifold and the Latin indices refer to the local Lorentz frame of the tangent space. The
most general Lagrangian of PGT is a quadratic function constructed
by the suitable scalar combinations of the irreducible decompositions of curvature and torsion. Here following Ref. \cite{Cemb} we
choose a Lagrangian in the form
\begin{eqnarray}
\label{action}
S \;= \; \frac{1}{16 \pi}\int d^4x \sqrt{-g}
\;\Bigl[
\mathcal{L}_{m}&-&R-\frac{1}{4}\left(d_{1}+d_{2}+4c_{1}+2c_{2}\right)\tilde{R}_{\lambda \rho \mu \nu}\tilde{R}^{\mu \nu \lambda \rho}
\\ \nonumber
&-&\frac{1}{4}\left(d_{1}+d_{2}\right)\tilde{R}^2
+c_{1}\tilde{R}_{\lambda \rho \mu \nu}\tilde{R}^{\lambda \rho \mu \nu}
+c_{2}\tilde{R}_{\lambda \rho \mu \nu}\tilde{R}^{\lambda \mu \rho \nu}
\\ [5pt] \nonumber
&+&d_{1}\tilde{R}_{\mu \nu}\tilde{R}^{\mu\nu}
+d_{2}\tilde{R}_{\mu\nu}\tilde{R}^{\nu\mu}
\Bigr].
\label{actioneq}
\end{eqnarray}
where $c_{1}, \;c_{2},\; d_{1}$ and $d_{2}$ are four constant parameters and
$\tilde{R}_{\mu \nu \lambda \rho}$ is the curvature tensor constructed from
the general non-symmetric connection; while $R$ refers to the Ricci scalar
constructed by the Levi-Civita connection. Note that with the use of the
identity
$\tilde{R}=R-2\nabla_{\lambda}T^{\rho
\lambda}\,_{\rho}+\frac{1}{4}T_{\lambda \mu \nu}T^{\lambda \mu
\nu}+\frac{1}{2}T_{\lambda \mu \nu}T^{\mu \lambda \nu}-T^{\mu}\,_{\mu
\lambda}T^{\nu}\,_{\nu}\,^{\lambda}$, one can rewrite the general PGT
Lagrangian with massless torsion in terms of the torsionless Einstein-Hilbert
Lagrangian \cite{Cemb}. By varying the above Lagrangian with respect to the
dynamical variables, \textit{i.e.} tetrad and spin connection, we get the general form
of the field equations in PGT. The two field equations can be succinctly
expressed in the following forms \cite{Shie}
\begin{eqnarray}
\label{FE1}
  \nabla_{\nu}H_{i}^{\mu\nu}-E_{i}^{~\mu}&=&{\cal T}_{i}^{~\mu}\;,\\
 \label{FE2}
  \nabla_{\nu}H_{ij}^{~~\mu\nu}-E_{ij}
  ^{~~\mu}&=&S_{ij}^{~~\mu}\; ,
 \end{eqnarray}
with the following definitions
 \begin{eqnarray}
 \label{H1}
  H_{i}^{~~\mu\nu}&:=&\frac{\partial e \mathcal{L}_{g}}{\partial\partial_{\nu} e_{\mu}^i}
  =2\frac{\partial e \mathcal{L}_{g}}{\partial T_{\nu\mu}{}^i}\;,\\ [7pt]
  \label{H2}
  H_{ij}{}^{\mu\nu}&:=&\frac{\partial e \mathcal{L}_{g}}{\partial\partial_{\nu}\Gamma_{\mu}^{~ij}}
  =2\frac{\partial e \mathcal{L}_{g}}{\partial R_{\nu\mu}{}^{ij}}\;,
 \end{eqnarray}
and
 \begin{eqnarray}
  E_{i} {}^{\mu}&:=&e^{\mu}{}_{i} e \mathcal{L}_{g}-T_{i \nu}{}^{j} H_{j} {}^{\nu\mu}
  -R_{i\nu}{}^{jk}H_{jk}{}^{\nu\mu}\;,\\
  E_{ij}{}^{\mu}&:=&H_{[ij]}{}^{\mu}\;,
 \end{eqnarray}

\noindent
where $\mathcal{L}_{g}$ is the gravitational Lagrangian included in Eq. \eqref{action}. The
source terms at the right hand side of Eqs. \eqref{FE1} and \eqref{FE2} are energy-momentum and spin density tensors respectively
and are defined by
\begin{eqnarray}
 {\cal T}_{i}{}^{\mu}&:=&\frac{\partial e\mathcal{L}_{m}}{\partial e_{\mu}{}^i}\;, \quad
S_{ij}{}^{\mu}:=
    \frac{\partial e\mathcal{L}_{m}}{\partial \Gamma_{\mu}{}^{ij}}
\end{eqnarray}
where $\mathcal{L}_{m}$ is the matter Lagrangian and $e$ is the determinant of
the tetrad.

Black hole solutions to the PGT field equations have been studied previously
by various authors \cite{Baekler1, Lee, Benn, McCrea1, Baekler2, McCrea2}. For example in Ref. \cite{Baekler3}, the authors found a solution
analogous to the Kerr solution of General Relativity. More recently  in Ref.
\cite{Cemb}, a new static spherically symmetric vacuum solution to the
Poincar{\'e} field equations \eqref{H1}  and \eqref{H2} for the Lagrangian in
the form of \eqref{action} has been found. This solution describes the exterior geometry
for a static spherically symmetric black hole with torsion. The solution
is analogous to the Reissner-Nordstr\"{o}m solution in General Relativity,
however here there is no specific electric charge. The solution can be
regarded as a modification of the Schwarzschild metric of General Relativity
where torsion provides extra terms in the metric. Here, we briefly review the
basic properties of this modified metric. The most general
 line element outside of a static, spherically symmetric black hole can be written as
\begin{equation}
\label{metric}
{ds}^2=-e^{\nu(r)}{dt}^2+e^{-\nu(r)}{dr}^2+r^2{d\theta}^2+r^2\sin^2\theta \; {d\phi}^2\;.
\end{equation}

In Riemann-Cartan geometry the torsion must also satisfy the intrinsic
symmetries of the background space-time. This means that in addition to the
metric,  the torsion tensor should also  satisfy the Killing equation
$L_{\xi} T_{\mu\nu}^{\rho}=0$ where $L_{\xi}$ is the Lie derivative in the
direction of $\xi$. Applying this Killing equation to static spherically symmetric space-time, the non-zero components of the torsion tensor can be explicitly written as
\cite{Cemb, Rauch2, Sur}
$${T^t}_{tr}=-{T^t}_{rt}=a(r)\,,\,{T^r}_{\theta\phi}=-{T^r}_{\phi\theta}=k(r)\sin{\theta} \; e^{\nu(r)}$$
$${T^r}_{tr}=-{T^r}_{rt}=a(r)e^{\nu(r)}\,,\,{T^t}_{\theta\phi}=-{T^t}_{\phi\theta}=k(r)\sin{\theta}$$
$${T^\phi}_{t\theta}=-{T^\phi}_{\theta
t}=\frac{h(r)e^{\nu(r)}}{\sin{\theta}}\,,\,{T^\theta}_{r\phi}=-{T^\theta}_{\phi
r}=h(r)\sin{\theta}$$ $${T^\phi}_{\theta
r}=-{T^\phi}_{r\theta}=\frac{h(r)}{\sin{\theta}}\,,\,{T^\theta}_{\phi
t}=-{T^\theta}_{t\phi}=h(r)\sin{\theta}\;e^{\nu(r)}$$ $${T^\theta}_{\theta
t}=-{T^\theta}_{t\theta}={T^\phi}_{\phi t}=-{T^\phi}_{t\phi}=g(r)e^{\nu(r)}$$
\begin{equation}\label{Tcomps} {T^\theta}_{r\theta}=-{T^\theta}_{\theta
r}={T^\phi}_{r\phi}=-{T^\phi}_{\phi r}=g(r) \end{equation}

\noindent
where $a(r),\;k(r),\;h(r)$ and $g(r)$ are four unknown functions to be determined by solving the field equations. Following Ref.
\cite{Cemb} the solution for the metric function $\nu(r)$ is
\begin{equation}
 \label{Sol}
 e^{\nu(r)} = \left( 1-{\frac{2m}{r}}+{\frac{s}{r^2}}\right)\;.
\end{equation}

Substituting this relation in the metric \eqref{metric}, we get a solution
that has the same symmetries as  the Schwarzschild metric of GR. The new metric has the form of the Reissner-Nordstr\"{o}m solution in general
relativity but without any electric charge as the source \cite{Cemb2}.  The parameters $m$
and $s$ are some constants of integration and can be related to the field strengths of curvature and
torsion, respectively. The torsion functions are also given by solving the
field equations \cite{Cemb}
\begin{equation}
  \label{TF}
a(r)=\frac{e^{\nu^\prime(r)}}{2e^{\nu(r)}}= \frac{(\frac{m}{r^2}-\frac{s}{r^3})}{(1-\frac{2m}{r}+\frac{s}{r^2})},
\quad
g(r)=-\frac{1}{2 r} ,
\quad
h(r)=-\frac{\sqrt{s}}{r e^{\nu(r)}},
\quad
k(r)=0\;.
\end{equation}

Let us briefly analyze the properties of the black hole solution given by Eqs.
\eqref{metric} and \eqref{Sol}. Here, the torsion function $s$ plays a role similar to the
electromagnetic charge in usual General relativistic Reissner-Nordstr\"{o}m
geometry. The  position of black hole horizons are given by
\begin{equation}
\label{horr}
R_\pm=m\pm\sqrt{m^2-s}\;,
\end{equation}
provided that the  following condition is satisfied
 \begin{equation}\label{conditionbh}
    m^2\geq s\;.
\end{equation}
The outer horizon $R_+$  can be regarded as the
Schwarzschild radius of the black hole.

\section{Gravitational lensing by a black hole with torsion in PGT}
\label{sec: 3}
Gravitational lensing by a black hole with torsion has been recently studied
in Ref. \cite{Zhang} in the framework of an extension to the
Einstein-Cartan-Sciama-Kibble (ECSK) theory presented in Ref. \cite{Shab}. In
that paper, the authors obtained static vacuum solutions by including
fourth-order scalar invariants constructed from curvature and torsion in the
ECSK Lagrangian and found both black hole solutions and naked singularities
in that setup. However, the presence of the fourth-order term in the
Lagrangian will add various complications to the gauge structure of the
theory. Here we choose the Poincar{\'e} gauge theory of gravity where the
Lagrangian is of the quadratic type given by \eqref{action} and study the
gravitational lensing by a black hole in this setup. Our aim is to determine
the deflection angle of light rays near a black hole, where the geometry of
the exterior space-time is described by metric \eqref{metric} and \eqref{Sol}
and torsion in the form of \eqref{Tcomps}. It should be noted that in general
Riemann-Cartan space-time, where the connection is not necessarily symmetric,
auto-parallel curves and metric geodesics do not coincide with each other. We
begin with the equation for null geodesics in this setup. Various components
of the metric geodesics, constructed by using the metric \eqref{metric} and
\eqref{Sol} are given by (we work in the equatorial plane $\theta=\pi/2$ for
simplicity, without any loss of generalization)

\begin{equation}
    \frac{d^2t}{dp^2}+2\Big(1-\frac{2m}{r}+\frac{s}{r^2}\Big)^{-1} \Big(\frac{m}{r^2}-\frac{s}{r^3}\Big) \frac{dr}{dp} \frac{dt}{dp}=0\;,
\end{equation}
\begin{eqnarray}
    \frac{d^2r}{dp^2}&-&\Big(1-\frac{2m}{r}+\frac{s}{r^2}\Big)^{-1} \Big(\frac{m}{r^2}-\frac{s}{r^3}\Big) \Big(\frac{dr}{dp}\Big)^2
    \\ \nonumber
    &+&\Big(\frac{m}{r^2}-\frac{s}{r^3}\Big) \Big(1-\frac{2m}{r}+\frac{s}{r^2}\Big)\Big(\frac{dt}{dp}\Big)^2
   -r \Big(1-\frac{2m}{r}+\frac{s}{r^2}\Big) \Big(\frac{d\phi}{dp}\Big)^2=0\;,
\end{eqnarray}
\begin{equation}
    \frac{d^2\phi}{dp^2}+ \frac{2}{r} \frac{dr}{dp}  \frac{d\phi}{dp}=0\;.
\end{equation}

Following the standard procedure outlined in Sect. $8.5$ of Ref. \cite{Weinberg}, the deflection angle for light rays near the black hole is given by the following integral
\begin{equation}
\label{geodef}
  \Phi= \int_{r_m}^{\infty}dr \frac{1}         {r\;\sqrt{{1-\frac{2m}{r}+\frac{s}{r^2}}} \sqrt{\frac{r^2\Big(1-\frac{2m}{r_m}+\frac{s}{r_m^2}\Big)}{r_m^2\Big(1-\frac{2m}{r}+\frac{s}{r^2}\Big)}-1}}\;,
\end{equation}
where ${r_m}$ is the distance of the closest approach. Note that although these equations are constructed by the Levi-Civita connection, the resulting deflection angle is still different from that of General Relativity as the components of the metric are now depend on the torsion parameter $s$.

\begin{figure}[tbp]
\centering
 \includegraphics[scale=0.55]{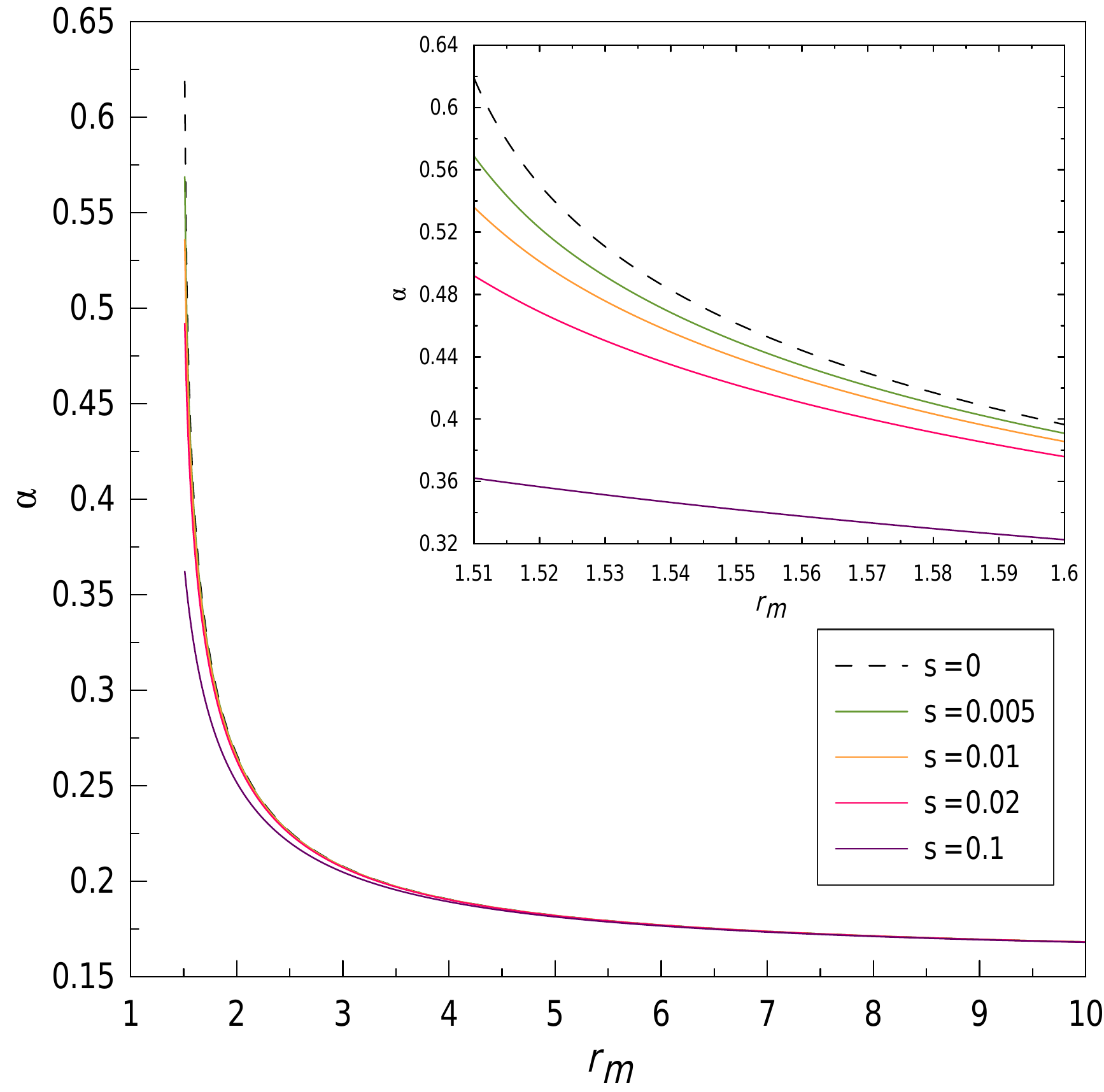}
\caption{Deflection angle defined as the integral in Eq. \eqref{geodef}
versus the minimum distance of the photons from the black hole $r_{m}$ for
different values of the torsion parameter $s$. In this
figure the source is assumed to be at infinity and
 $m=0.5$. The horizontal axis is in terms of the Schwarzschild radius.}
 \label{fig: Geoinf}
 \end{figure}

On the other hand, various components of the auto-parallel equation constructed by the full connection given in the \ref{App: A} are

\begin{eqnarray}
\label{auto1}
\frac{d^2t}{dp^2}&+&\left( \nu^{\prime}(r)+a(r)\right)
\frac{dt}{dp} {\frac{dr}{dp}} - a(r) e^{-\nu(r)} \left(
{\frac{dr}{dp}}\right) ^2
+r^2 g(r) \left(
{\frac{d\theta}{dp}}\right) ^2
\\ [5pt] \nonumber
&+&\ r^2 \sin^2\theta \; g(r)
  \left( {\frac{d\phi}{dp}}\right)^2=0\;,
\end{eqnarray}

\begin{eqnarray}
\label{auto2}
\frac{d^2r}{dp^2}&-&\frac{1}{2} \nu^\prime(r)
\left( \frac{dr}{dp}\right) ^2
-\left( a(r)
e^{\nu(r)} \right) \frac{dt}{dp} \frac{dr}{dp}
+\left( \frac{1}{2} \nu^\prime (r) + a(r)
\right) e^{2\nu(r)} \left(\frac{dt}{dp} \right) ^2
\\ \nonumber
 &+&r e^{\nu(r)} \left(r
g(r) -1 \right) \left( \frac{d\theta}{dp}\right) ^2
+r e^{\nu(r)} \sin^2\theta \: \left(r g(r) -1 \right)
\left( \frac{d\phi}{dp}\right) ^2=0\;,
\\ [20pt]
\label{auto3}
\frac{d^2\theta}{dp^2} &-& \sin\theta \: e^{\nu(r)}
\left(\frac{k(r)}{r^2} + h(r) \right) \frac{dt}{dp}
\frac{d\phi}{dp}
+\left( \frac{2}{r} - g(r) \right)
\frac{dr}{dp} \frac{d\theta}{dp}
\\ \nonumber
&+&\sin\theta \left( \frac{k(r)}{r^2} + h(r)\right)
 \;\frac{d\phi}{dp} \frac{dr}{dp}
-
\sin\theta \;\cos\theta \left( \frac{d\phi}{dp}\right)^2
+\left(g(r) e^{\nu(r)}\right)  \frac{dt}{dp}  \frac{d\theta}{dp}=0\;,
\\ [25pt]
\label{auto4}
\frac{d^2\phi}{dp^2}&+&\left(\frac{2}{r} - g(r) \right) \frac{dr}{dp}
\frac{d\phi}{dp}+\left(g(r) e^{\nu(r)} \right)  \frac{dt}{dp}
\frac{d\phi}{dp}
\\ \nonumber
&+&\frac{e^{\nu(r)}}{\sin\theta}
 \left( \frac{k(r)}{r^2} + h(r)\right) \frac{dt}{dp}
\frac{d\theta}{dp}
+2\left(\cot{\theta} \right)
\frac{d\phi}{dp} \frac{d\theta}{dp} -\frac{1}{\sin{\theta}}
\left(k(r) + h(r) \right)  \frac{dr}{dp} \frac{d\theta}{dp}=0\;.
\end{eqnarray}
\;
\newline
Again, we work in the equatorial plane $\theta=\pi/2$. In this case,
Eq. \eqref{auto3} gives
\begin{equation}
\label{equat}
 e^{\nu(r)} \frac{dt}{dp} = \frac{dr}{dp}\;.
\end{equation}

\noindent
Using the above, Eqs. \eqref{auto1}, \eqref{auto2} and \eqref{auto4} can be written
as
\quad
\begin{equation}
\label{auto11}
 \frac{d^2t}{dp}+\left(\nu^\prime(r)  e^{\nu(r)} \right)
\left( \frac{dt}{dp}\right)^2+\left( r^2 g(r)\right) \left(
\frac{d\phi}{dp}\right) ^2=0\;,
\end{equation}
\begin{equation}
\label{auto22}
\frac{d^2r}{dp^2}+\left(r e^{\nu(r)}   \left(r g(r) -1 \right)  \right)
\left(\frac{d\phi}{dp} \right) ^2=0\;,
\end{equation}
\begin{equation}
\label{auto44}
\frac{d^2\phi}{dp^2}+ \frac{2}{r} \frac{dr}{dp}  \frac{d\phi}{dp}=0\;.
\end{equation}

\noindent
Using torsion functions \eqref{TF}, the Eqs.  \eqref{auto11} and \eqref{auto22}  become
\begin{equation}
\label{auto111}
\frac{d^2t}{dp^2}+2\left(\frac{m}{r^2} - \frac{s}{r^3}
\right) \left( \frac{dt}{dp}\right) ^2 - \frac{r}{2} \left(
\frac{d\phi}{dp}\right) ^2=0\;,
\end{equation}
\begin{equation}
\label{auto222}
\frac{d^2r}{dp^2} - \frac{3}{2} r \left( 1-\frac{2m}{r}+ \frac{s}{r^2}\right)
\left( \frac{d\phi}{dp}\right) ^2=0\;.
\end{equation}

\noindent
Eq. \eqref{auto44} can be rewritten as
\begin{equation}
\frac{d}{dp} \left\lbrace \ln{\frac{d\phi}{dp}}+\ln{r^2}\right\rbrace =0\;,
\end{equation}
which immediately gives
\begin{equation}
\label{J}
r^2
\frac{d\phi}{dp}=J\;,
\end{equation}
where $J$ is a constant of motion interpreted as the angular momentum. We also have
\begin{equation}
\label{AM}
ds^2=E
\,
dp^2\;.
\end{equation}
Using Eqs. \eqref{equat}, \eqref{auto111} and \eqref{J}  to find $d\phi/dp$,
$(dt/dp)^{2}$ and ${d^{2}t}/{dp^{2}}$ respectively;
noting that $E=0$ for photons, Eq.  \eqref{auto222} takes the following form
after some simplification
\begin{eqnarray}
\label{eqE}
\frac{d^{2}r}{dp^{2}}&+&2\left(\frac{J}{r}\right)^{2} \left[\left(\frac{m}{r^{2}}-\frac{s}{r^{3}}\right)-\frac{1}{4r} \left(1-\frac{2m}{r}+\frac{s}{r^{2}}\right)\right] =0\;.
\end{eqnarray}

We are interested in the trajectory of the photons in the $r-\phi$ plane. To
find $r(\phi)$, we can eliminate $dp$ from Eq. \eqref{eqE} by noting that
\begin{eqnarray}
\frac{d^{2}r}{dp^{2}}= \frac{d}{dp}\frac{dr}{dp}
     &=&\frac{d\phi}{dp} \frac{d}{d\phi}\left(\frac{d\phi}{dp} \frac{dr}{d\phi}\right) \nonumber
     \\ [4pt]
     &=&\left(\frac{J}{r^{2}}\right)^{2} \left[\frac{d^{2}r}{d\phi^{2}}-\frac{2}{r}\left(\frac{dr}{d\phi}\right)^{2}\right]\;.
\end{eqnarray}
Using this, Eq. \eqref{eqE} takes the following form
\begin{eqnarray}
\label{eqEE}
\frac{d^{2}r}{d\phi^{2}}-\frac{2}{r} \left(\frac{dr}{d\phi}\right)^2&+&2r^{2} \left(\frac{m}{r^{2}}-\frac{s}{r^{3}}\right)
-\frac{r}{2} \left(1-\frac{2m}{r}+\frac{s}{r^{2}}\right)=0\;.
\end{eqnarray}

\noindent
This equation immediately gives the following integral for the deflection angle
\begin{equation}
\label{DefAng}
\Phi=\pm\int^{r_S}_{r_m}\frac{2\:dr}{\sqrt{4r^{4}C_{1}-2r^{2}+8mr-5s}}-C_{2}\;,
\end{equation}
where $C_{1}$ and $C_{2}$ are constants of integration and $r_{m}$ and
$r_{S}$ are the closest approach distance and position of the source,
respectively. The integral in Eq. \eqref{DefAng} is of the elliptic type which has no analytic solutions and should
be solved numerically. The numerical analysis was performed using the Monte Carlo method to find the deflection angle when the
geometry outside of the black hole lens is given by the metric of the Eqs. \eqref{metric} and
\eqref{Sol} where also a torsion function in the form of Eq. \eqref{Tcomps} is present. We also
employed the Nintegrate package in the Mathematica software system to cross-check our results.

\subsection{Deflection angle}
The deflection angle when light rays follow the metric geodesics is given by
Eq. \eqref{geodef}. Figure [\ref{fig: Geoinf}] shows the deflection angle in
terms of the minimum distance $r_m$ for different values of the torsion
parameter $s$. The value of the mass parameter $m$ is chosen to be $m=0.5$ in
this figure and the minimum distance $r_m$ in the horizontal axis is in terms
of the Schwarzschild radius of the black hole.  The plot is obtained by
solving the integral of Eq. \eqref{geodef} numerically using the Monte Carlo
method and shows a slight change in the deflection angle when the torsion
parameter $s$ is increased. This in turn leads to a slight change to the
position of the observed images compared to what is expected from GR. It is obvious from the figure that the value of $\Phi$ in Eq.
\eqref{geodef} decreases when the torsion parameter $s$ increases.

\begin{figure}[tbp]
\centering
 \includegraphics[scale=0.55]{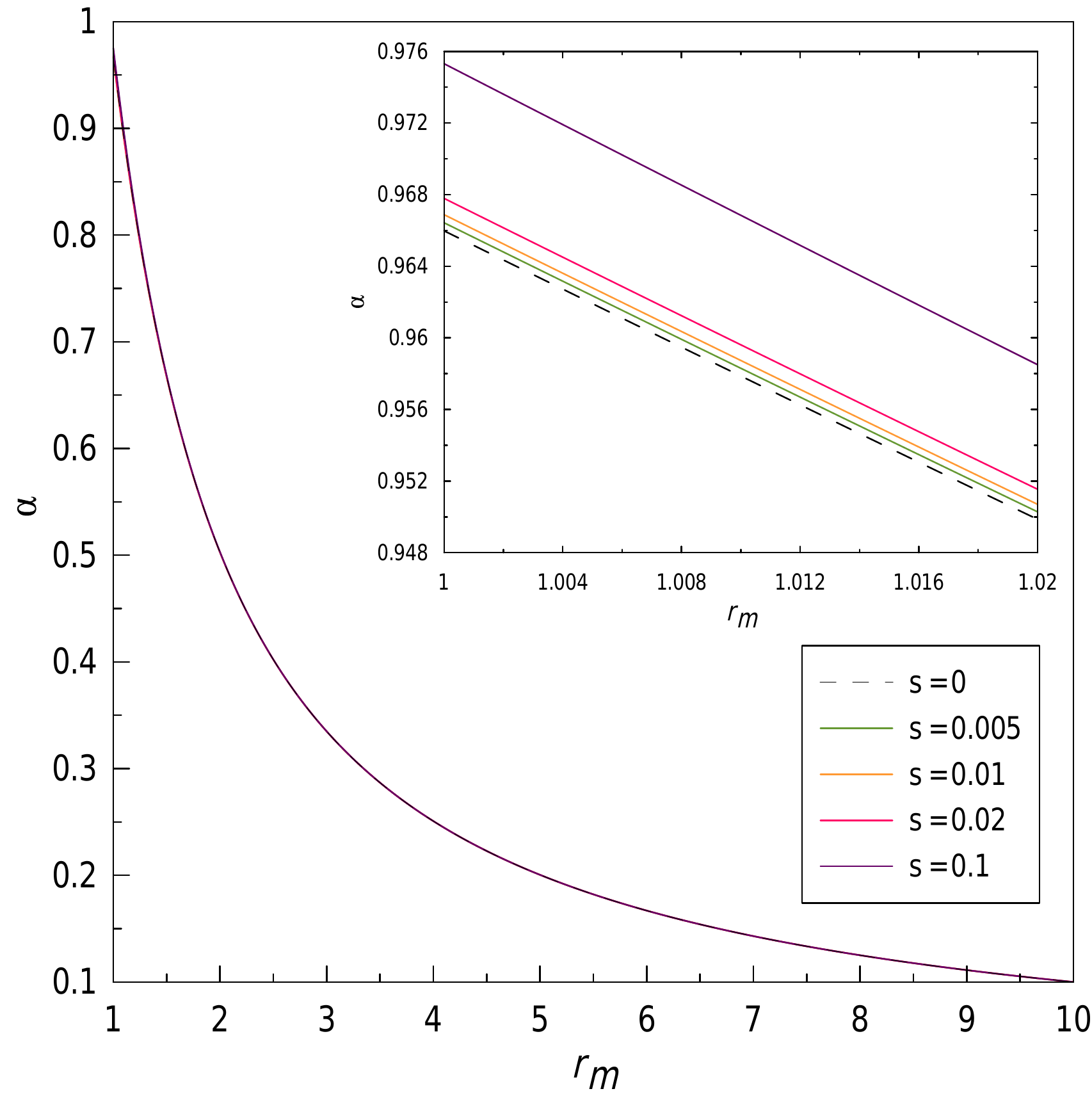}
\caption{Deflection angle defined as the integral in Eq. \eqref{DefAng}
versus the minimum distance of the photons from the black hole $r_{m}$ for
different values of the torsion parameter $s$ in Eq. \eqref{Sol}. In this
figure the source is assumed to be at infinity \textit{i.e.} very far from
the lens. The values of other constants are $m=0.5$ and $C_{1}=1$. The horizontal axis is in terms of the Schwarzschild radius.}
\label{fig: infi}
\end{figure}

In the case of the auto-parallel curves, first let us assume that the light
ray comes from infinity ($r_S=\infty$) and reaches a minimum distance $r_{m}$
from the lens, before reaching the observer. Figure [\ref{fig: infi}] shows
the deflection angle in terms of arcseconds with respect to $r_{m}$ for
different values of the torsion parameter $s$. The values of other constants
are set to $C_{1}=1 \,,\, m=0.5$.  As expected, the deflection angle decreases with increasing minimum distance $r_{m}$. Also, the
effects of torsion only become noticeable for small values of $r_{m}$ and
generally the deviation from GR increases with increasing the
absolute value of the torsion parameter $s$.

\begin{figure*}[tbp]
\centering
 \includegraphics[scale=0.55]{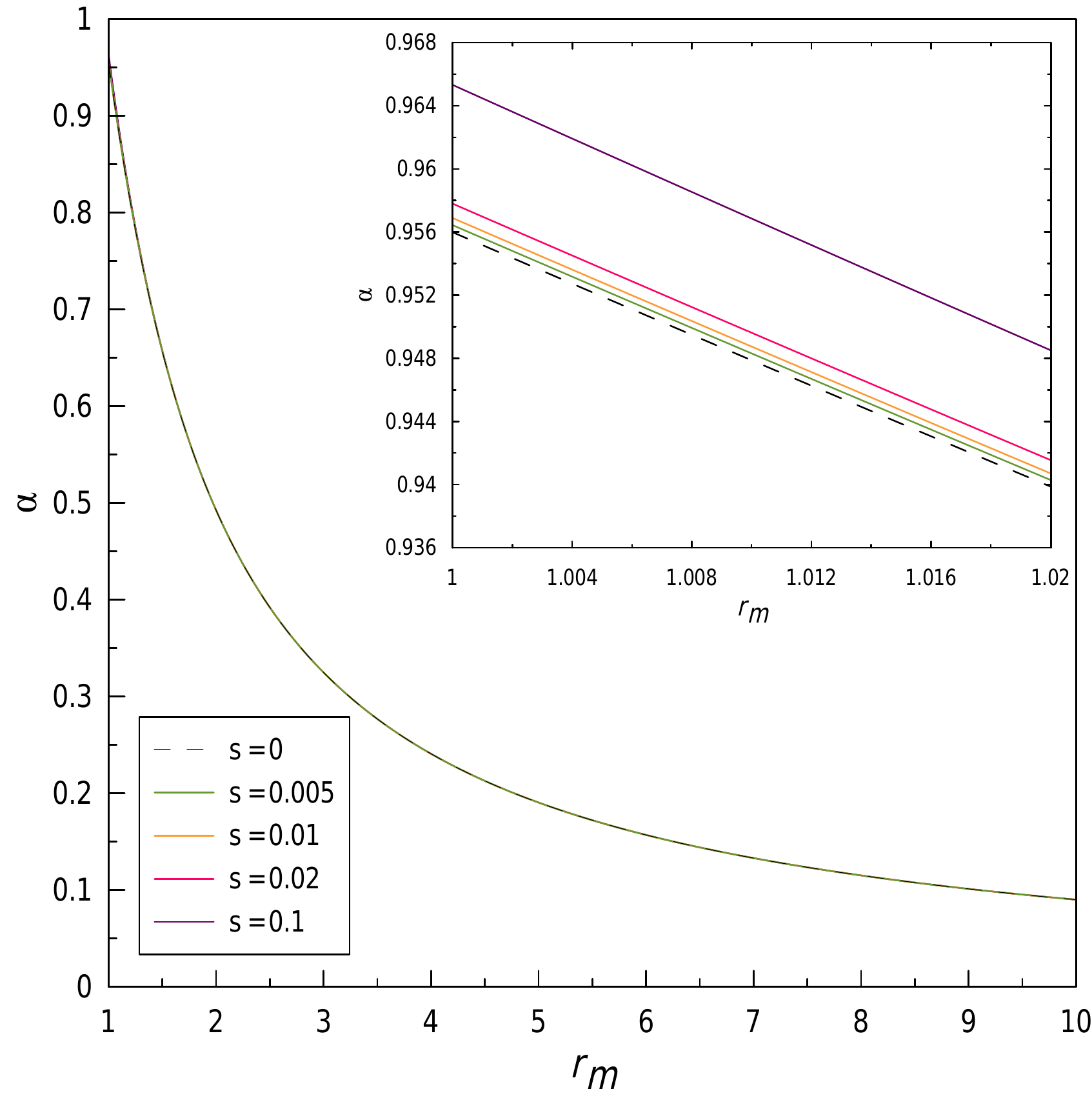}
\caption{Deflection angle defined as the integral in Eq. \eqref{DefAng}
versus the minimum distance of the photons from the black hole $r_{m}$ for
different values of the torsion parameter $s$ in Eq. \eqref{Sol}. In this
figure the source is assumed to be at $r_{S}=100$. The values of other
constants are $m=0.5$ and $C_{1}=1$. The horizontal axis is in terms of the Schwarzschild radius.}
\label{fig: 100}
\end{figure*}

\begin{figure*}[thbp]
\begin{tabular}{rl}
\includegraphics[width=7cm]{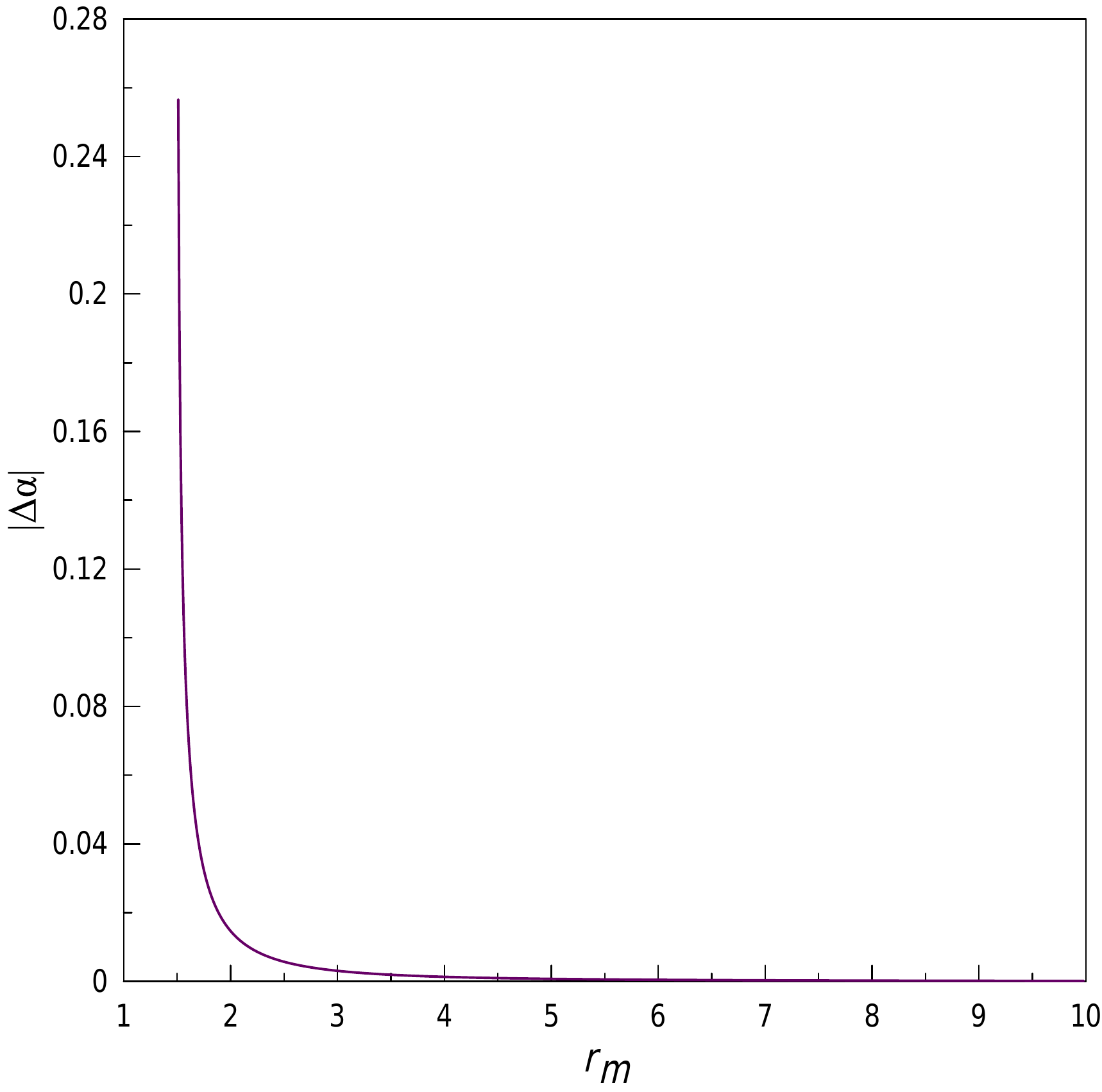}&
\includegraphics[width=7cm]{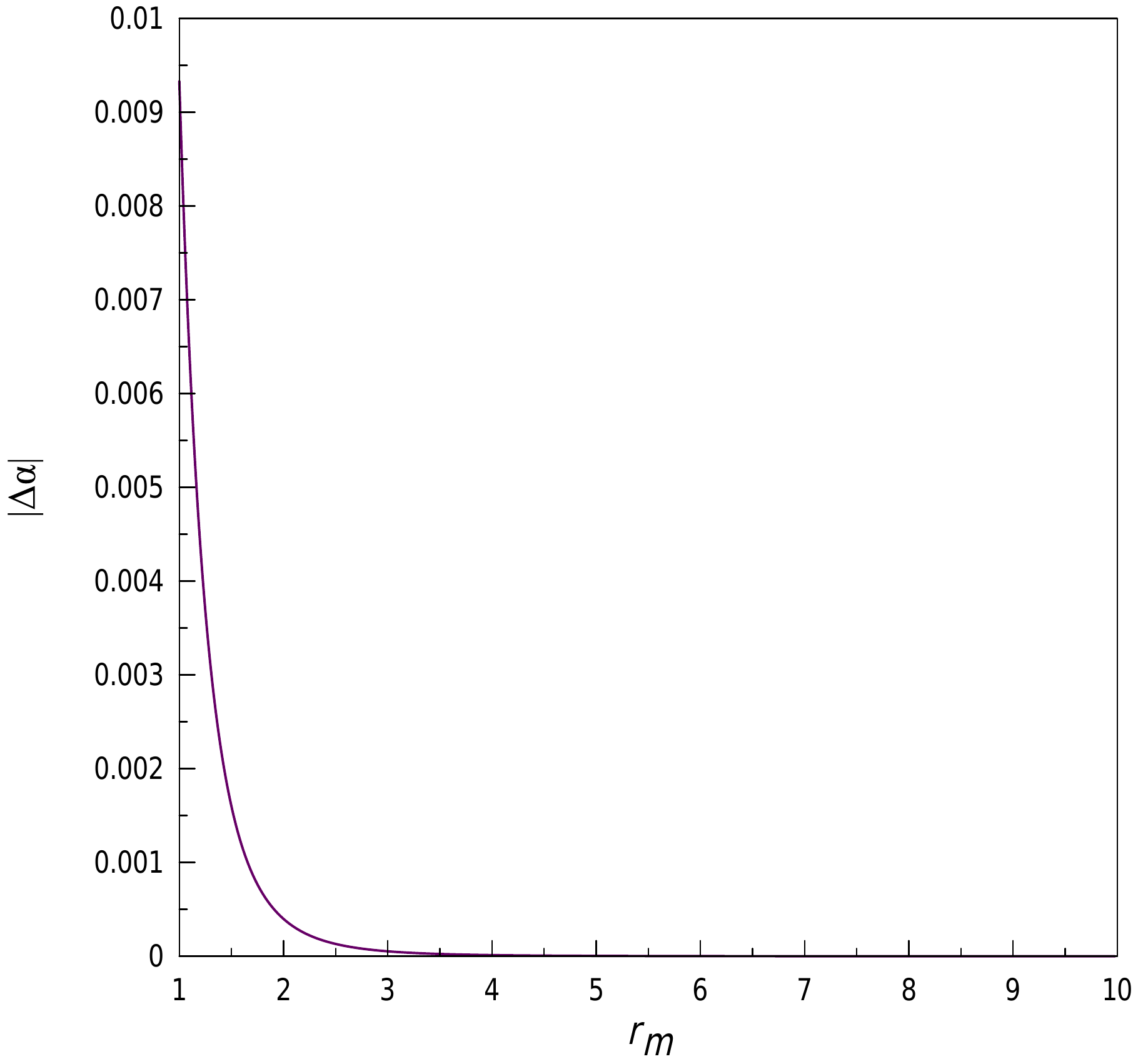}\\
\end{tabular}
\caption{\small{The difference between general relativistic deflection angle and the black hole with torsion for metric geodesics (left) and auto-parallel curves (right) for the case of the torsion parameter $s=0.1$. The difference is in the order of microarcseconds and becomes greater when the minimum distance $r_m$ gets closer to the black hole horizon. The vertical axes in the figures are in terms of milliarcseconds (mas).}}
\label{fig: diff}
\end{figure*}

For a more realistic system, we next assume that $r_{S}$ has a finite value.
In Figure [\ref{fig: 100}], we plot the deflection angle versus the minimum
distance $r_{m}$ for $r_{S}=100$ and different values of torsion parameter
$s$. The values of other constants are as before. The figure again shows the
effects of torsion on the deflection angle for light rays lensed by a black
hole in non-Riemannian space-times. For comparison, the case $s=0$ (General
relativistic result) is also included in all of the figures. Generally, in
the case of the auto-parallel curves, positive values of the torsion
parameter $s$ tend to increase the deflection angle compared to the General
relativistic case ($s=0$) while negative values of $s$ will decrease the
deflection angle.
Figure [\ref{fig: diff}] shows the difference between general relativistic
deflection angle and PGT black hole described by metric \eqref{Sol} for the
case $s=0.1$. The left and right figures are for metric geodesics deflection
angle given by \eqref{geodef} and auto-parallel case given by \eqref{DefAng}
respectively. The difference, generally increases when the minimum distance
$r_m$ approaches the black hole horizon. The expected correction from the effects
of torsion is in the order of microarcseconds. Similar estimations for the
correction of deflection angle were obtained in Refs.
\cite{Cappo, Horv}  in the weak field limit of fourth order
$f(R)$ gravity.


\subsection{Position of the images}
\begin{figure*}[tbp]
\centering
 \includegraphics[scale=0.55]{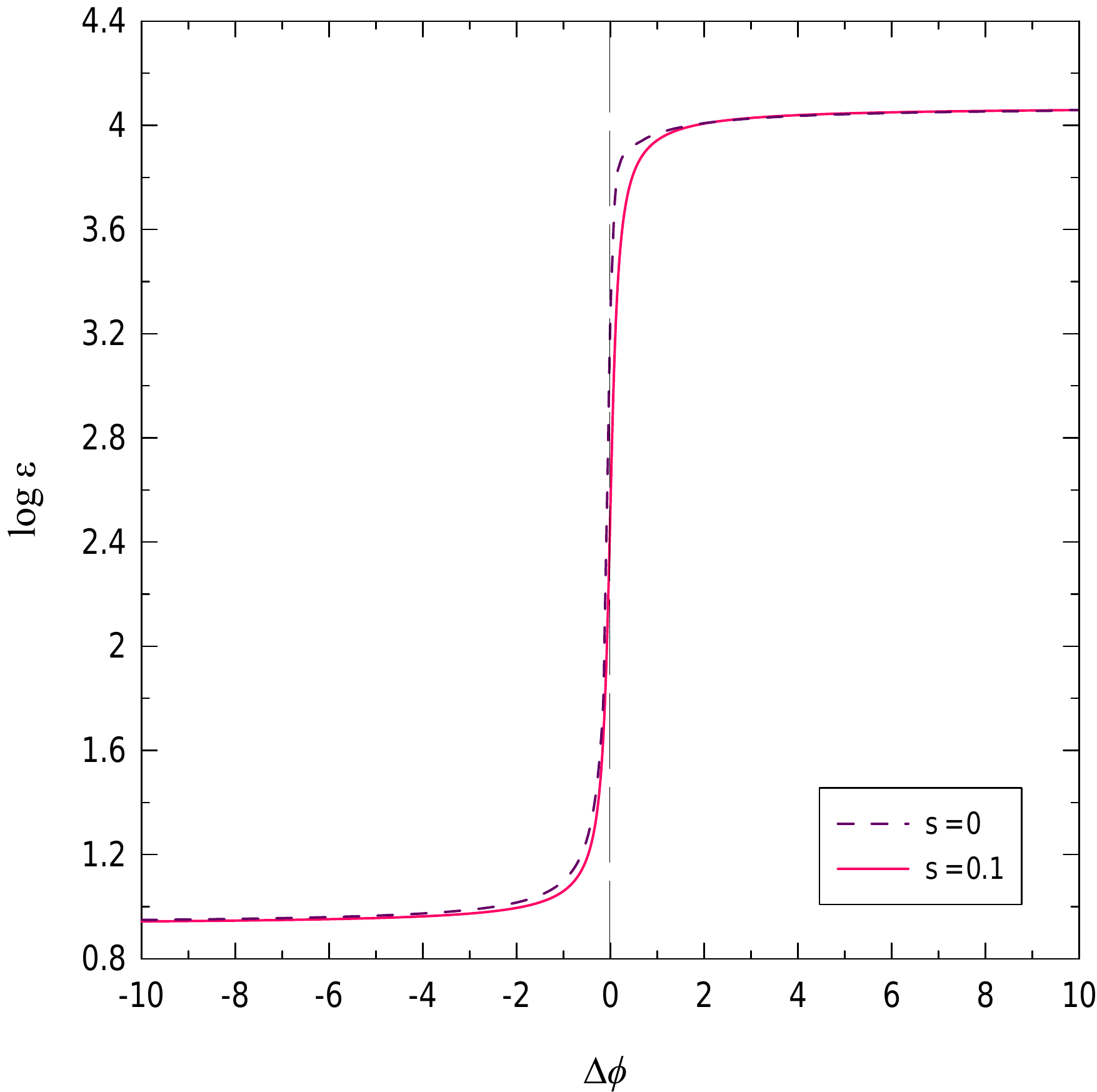}
\caption{Logarithmic behaviour of the position of images versus the
position of the source $\Delta\phi$ for torsion parameter $s=0.1$ (solid line)
and $s=0$ (dashed line). Vertical axis shows $\log\epsilon$ with
$\epsilon\equiv\left(\frac{\theta}{\bar{\theta}}-1 \right)$ where $\theta$ is
the position of the image and $\bar{\theta}$ is the position of the shadow of
the black hole. Both $\theta$ and $\bar{\theta}$ are in microarcseconds ($\mu
as$) while $\Delta\phi$ is in milliarcseconds ($mas$).  The source and the
observer assumed to be very far from the lens.}
\label{fig: Position}
\end{figure*}
The position of images can be obtained by utilizing the black hole lens
equation which can be written as \cite{Bozza3}
\begin{equation}
\Phi(u,r_{O},r_{S})=\Delta \phi \equiv  \phi_O -\phi_S +2n\pi \;.
\label{ExactLE}
\end{equation}
Here $n$ is an integer number, $\Phi$ is given by Eq. \eqref{geodef} for the
metric geodesics and Eq. \eqref{DefAng} for the auto-parallel curves; $\Delta
\phi$ is the relative azimuthal position between the source and the observer
and  $ \phi_O$ and $\phi_S$ are the azimuthal position of the observer and
the source, respectively. Generally, the position of the images can be
obtained by finding the impact parameter $u$ from the above lens equation and
use it to find the angle in which the observer detects the photon; however,
there is a situation where this equation can be simplified greatly. When the
source and the observer are both very far from the lens, we can use the
following approximation
\begin{equation}
\sin\theta\simeq \frac{u}{r_{O}}\quad\quad\,,\quad\quad \sin\theta_{S}=\frac{u}{r_{S}}
\end{equation}
to eliminate $r_{O}$ and $r_{S}$ from Eq. \eqref{ExactLE} and get the Ohanian lens equation \cite{Ohan}
\begin{equation}
\alpha-\theta-\theta_S=-\gamma\equiv\Delta \phi-\pi\;.
\label{LensEQOha}
\end{equation}
Here $\alpha$ is the deflection angle between asymptotic incoming and
outgoing trajectories of the photon \textit{i.e.}
$\alpha=\Phi(u,\infty,\infty)-\pi$ and $\theta$ is the detection angle of the
photon by the observer. $\theta_S$ can be related to $\theta$ by the relation
\begin{equation}
\theta_S=\sin^{-1}\Big(\frac{d_{OL}}{d_{LS}}\sin\theta\Big)\;,
\end{equation}
where $d_{OL}$ and $d_{LS}$ are distances between observer and the lens, and
lens and the source respectively.

Figure [\ref{fig: Position}] shows the position of images for different values of the source
position. Here we assumed the source and the observer to be very far from the
lens, so the Ohanian lens equation would be a reasonable approximation.  The
vertical axis is $\log\epsilon$ where $\epsilon$ is defined as
\begin{equation}
\epsilon\equiv\left(\frac{\theta}{\bar{\theta}}-1 \right)\;,
\end{equation}
Where $\bar{\theta}$ is the position of the photon sphere (the so-called
shadow of the black hole). The solid line in Figure [\ref{fig: Position}] shows the position of
images for the value of $s=0.1$ for the torsion parameter in Eq. \eqref{Sol}.
For comparison again we plot the position of images for the torsion-less case
$s=0$. The figure shows that the effects of torsion slightly alter the
position of images compared to the General relativistic case, specially when the
source and the lens are slightly misaligned. Note that $\Delta\phi=0$ and
$\Delta\phi=\pi$ indicate the source in front of the black hole and directly
behind the black hole, respectively. For positive values of $\Delta\phi$ we
have primary images where the image is on the same side of the source while
for negative values of $\Delta\phi$ (secondary images), the image is located
at the opposite side of the source. Note that for both of these images we
have $n=0$ in Eq. \eqref{ExactLE}. Higher order images can be obtained by choosing
different values of $n$ in the lens equation. Generally, as can be seen from
the figure, by increasing the value of $\Delta\phi$, primary images
move to the position of the source, while the secondary image moves closer
to the shadow of the black hole. In case of the perfect alignment between
source, lens and the observers, the primary and secondary images merge to produce an Einstein's ring.

\section{Conclusion}
\label{sec: 4}
In this paper, we study gravitational lensing in the framework of a new
static spherically symmetric black hole solution recently found in the
framework of the Poincar{\'e} gauge theory of gravity. In this solution, the
effects of torsion appear as a single parameter in the line element. The
resulting metric is of the Reissner-Nordstr\"{o}m type, but with a dynamical
torsion instead of an electric or magnetic source.

In the Poincar{\'e} gauge theory of gravity, torsion couples to the spin of
the matter and can alter the path of particles moving through the
gravitational field. This effect on the orbit of particles can be more
pronounced in the extremely strong gravitational field of a black hole, or
when the spin density of the matter (usually defined as the expectation value
of the square of the spin density tensor) is very high. The effect of
dynamical torsion on the equation of motion of particles is two-fold. For
particles with non-vanishing spin, the coupling of torsion to the spin
density tensor results in the equation of motion different from the geodesic
equations of GR. However, for spinless particles, the
equations of motion reduce to that of GR. On the other hand,
even for particles with vanishing spin, torsion can alter the solution to the
field equations, \textit{i.e.} the tetrad field and as a result, the path of
particles and light rays will be different from their corresponding
trajectories in General Relativity. It has been shown previously that in the
absence of an explicit coupling between the electromagnetic field and torsion
in the Lagrangian, consistency with the Maxwell's equations compels photons
to follow the metric geodesics of the space-time \cite{Puetz, Yasskin, Nom}.
However, even in this case the deflection angle and the position of images
will not be the same as in GR, as the metric around the black
hole will be modified.

The effects of the torsion parameter in Eq. \eqref{Sol} on the gravitational
lensing near a black hole are summarized in Figs. [1-4]. Here, both
deflection angle and position of images slightly differ from the General
relativistic case, specially when the minimum distance of the incoming photon
from the black hole is small \textit{i.e.} near the photon sphere. This may
provide a way to detect dynamical torsion and test various modified gravity
theories by observing the gravitational lensing by an extremely strong
gravitational field of black holes or other extremely dense objects. This is
also may change the estimation of the Hubble parameter from such a lensing
system and may help in the resolution of the so-called Hubble tension
problem.

Finally, we briefly discuss the possibility of observing the effects
of torsion on the black hole lensing parameters. As the magnitude of
corrections is directly related to the ratio of the black hole mass and its
distance to the observer, the best candidates for observing the effects and
constraining the torsion parameter seems to be the supermassive black holes
at the center of the milky way and nearby galaxies (specially M87).
Gravitational lensing of the stars rotating the black hole at the center of
the milky way with eccentric orbits is quite common \cite{Bozza4}. One of the most
interesting lensing candidates is the star $S6$, because the eccentricity of
its orbit brings it quite close to the black hole horizon \cite{Bozza4}.  In
this range any possible effects of torsion are expected to be the largest. Note
that separation of images for this and similar stars are at the limits of
current observational capabilities; nonetheless, these systems provide a
convenient way to test various modified gravity theories via gravitational
lensing.

\appendix
\section{Components of Affine Connection}
\label{App: A}
Here, we present the explicit components of the affine connection satisfying the symmetries of a static spherically symmetric space-time with torsion
\begin{alignat*}{2}
\Gamma^0_{01}&=-\Gamma^1_{11}=\frac{1}{2}\nu^\prime(r), \;\qquad && \Gamma^0_{10}= \frac{1}{2}\nu^\prime(r)+a(r),\;\nonumber
\\[5pt]
\Gamma^0_{11}&=-a(r) e^{-\nu(r)},\;\qquad && \Gamma^0_{22}=r^2 g(r),\;  \nonumber
\\[5pt]
\Gamma^0_{32}&=-\Gamma^0_{23}=\frac{1}{2} k(r) \sin\theta, \;\qquad && \Gamma^0_{33}=r^2 \sin^2{\theta}\;g(r),\;\nonumber
\\[5pt]
\Gamma^1_{00}&=\left( \frac{1}{2} \nu^\prime(r) + a(r)\right) e^{2\nu(r)}, \;\qquad &&\Gamma^1_{01}= -a(r)e^{\nu(r)}, \; \nonumber
\\[5pt]
\Gamma^1_{22}&= r e^{\nu(r)} (rg(r)-1),\; \qquad &&
\Gamma^1_{32}=-\Gamma^1_{23}= \frac{1}{2} k(r) \sin\theta \; e^{\nu(r)},\;\nonumber
\\[5pt]
\Gamma^1_{33}&=r \sin^2\theta \; e^{\nu(r)} (r g(r)-1),\; \qquad && \Gamma^2_{02}= g(r)e^{\nu(r)},\;\nonumber
\\[5pt]
\Gamma^2_{03}&=-\frac{1}{2r^2} k(r) \sin\theta \; e^{\nu(r)}, \; \qquad && \Gamma^2_{12}=\Gamma^3_{13}= \frac{1}{r}-g(r),\;\nonumber
\\[5pt]
\Gamma^2_{13}&=\frac{1}{2r^2} k(r)\sin\theta, \; \qquad && \Gamma^2_{21}=\Gamma^3_{31}= \frac{1}{r},\;\nonumber
\\[5pt]
\Gamma^2_{30}&=-\frac{1}{2r^2}\left(2r^2 h(r)+ k(r) \right) \sin\theta \;e^{\nu(r)},\;\qquad && \Gamma^2_{31}=\frac{1}{2r^2} \left(2 r^2 h(r)+ k(r) \right)\;\sin\theta,\; \nonumber
\\[5pt]
\Gamma^2_{33}&=-\sin\theta \:\cos\theta, \qquad && \Gamma^3_{03}= g(r) e^{\nu(r)}, \; \nonumber
\\[5pt]
\Gamma^3_{02}&=\frac{1}{2} \frac{k(r) e^{\nu(r)}}{r^2 \; \sin\theta},\;\qquad && \Gamma^3_{12}=-\frac{1}{2} \frac{k(r)}{r^2\;\sin\theta},\; \nonumber
\\[10pt]
\Gamma^3_{20}&=\frac{1}{2}\frac{\left(2 r^2 h(r)+k(r) \right) e^{\nu(r)} }{r^2 \sin\theta},\; \qquad &&
\Gamma^3_{21}=- \frac{1}{2} \frac{\left( 2 r^2 h(r) + k(r)\right) }{r^2 \sin\theta},\; \nonumber
\\[5pt]
\Gamma^3_{23}&= \Gamma^3_{32}=\cot\theta,
\end{alignat*}
Where functions $a(r),\; k(r),\; h(r)$ and $g(r)$ are given by Eq. \eqref{TF} and we have

\begin{equation}
b(r)=\frac{e^{\nu^\prime(r)}}{2},
\qquad
c(r)=\frac{e^{\nu(r)}}{2r},
\qquad
d(r)=\frac{\sqrt{s}}{r},
\qquad
l(r)=0.
\end{equation}

\end{document}